\documentclass{PoS}
\usepackage{float}
\usepackage[caption= false]{subfig}
\usepackage[]{graphicx}
\usepackage[utf8x]{inputenc} 
\usepackage{xspace}
\title{The Onset of Hadronic Supercriticality in Expanding Sources}

\ShortTitle{Hadronic Supercriticality in expanding sources}

\author{{Ioulia Florou}\\
       Department of Physics, National \& Kapodistrian University of Athens\\
        E-mail: \email{iflorou@phys.uoa.gr}}
\author{Apostolos Mastichiadis\\
       Department of Physics, National \& Kapodistrian University of Athens\\
        E-mail: \email{amastich@phys.uoa.gr}}

\abstract{An overlooked property of hadronic models is that they can become supercritical by abruptly transforming the energy stored in the relativistic protons into radiation. Supercriticality manifests itself when the proton density exceeds a critical value. We seek to map the complete parameter space of this behaviour in cases where a source, consisting of relativistic protons, is expanding so adiabatic losses become important. We search those critical values that lead the system to the supercritical regime and are closely related to Gamma Ray Bursts. For this reason we adopt the Relativistic Blast Wave (RBW) model which is thought to describe the production of a GRB after the initial explosion }

\FullConference{High Energy Phenomena in Relativistic Outflows VII - HEPRO VII\\
		9-12 July 2019\\
		Facultat de Física, Universitat de Barcelona, Spain}

\begin{document}
\section{Hadronic Supercriticalities in GRBs}

The hadronic models for compact non-thermal astrophysical sources, such as Gamma Ray Bursts (GRBs), are a viable alternative to the more popular leptonic ones \cite{DA03}, \cite{Asano2007}. Apart from \mbox having implications for the sites of Cosmic Ray acceleration and for high energy neutrino astronomy \cite{W95}, which start becoming testable by ongoing observations \cite{Icecube18}, hadronic models have an intriguing property, coined as supercriticality \cite{Kazanas02}. Relativistic protons become supercritical when their energy density inside a source exceeds a certain  threshold \cite{KM92}. In that case, the energy stored into protons is abruptly released as non thermal radiation and the system is led to a high efficiency regime that is non linear. As a consequence, photons increase exponentially, leading to the onset of  flares which could be analogous to those observed from GRBs.

The required critical conditions for the onset of hadronic supercriticality to GRBs have been investigated analytically in previous works \cite{14MPGM}, \cite{2018PM}. It was shown that for parameters relevant to such phenomena, the system produces photon luminosities and spectra similar to those observed during the GRB prompt emission phase. However in all previous efforts it was assumed that the source is stationary, i.e. non-expanding. One interesting  question arises if we consider that the source does expand as, in such a case, the dynamical evolution of the source will alter the conditions for supercriticality. In this work, we examine such a scenario and make a first application of this model to GRB emission.

\section{Numerical Approach }

In order to examine the properties of the hadronic supercriticality in a stationary case, the standard framework of an one-zone radiation model is adopted, i.e pre-accelerated relativistic protons are injected uniformly  inside a spherical source, containing a tangled magnetic field, and produce radiation via various processes (synchrotron and photohadronic interactions). The hadronic system consists of three stable populations, protons, electrons and photons.  The hadronic supercriticality manifests itself when the proton energy density exceeds a critical limit in which case relativistic protons become targets for their own radiation and cool abruptly.

In order to examine whether an expanding source still shows supercritical behaviour, one has to adopt a different framework to that of previous works. Here we consider that the spherical volume, inside which relativistic protons are injected, is expanding with a constant expansion velocity $u_{exp}$. 
 
 The radius of the source is changing according to the relation: 
$R=R_{0}+u_{exp}t$
where t is the comoving time and $R_{0}$ the initial radius. For the scope of our preliminary analysis we assume that the magnetic field inside the source is either constant or varies as $B \sim R^{-1}$.

The evolution of these species (protons, electrons and photons) can be fully described by three coupled integro-differential kinetic equations \cite{DMPM12}. The injection and loss terms used in the kinetic equations include the following processes: synchrotron radiation for both electron and protons, photopair and photopion production, photon-photon pair production, inverse Compton scattering, synchrotron self absorption, pair annihilation, photon downscattering on cold electrons. Since the source is expanding the adiabatic losses are also taken into account \cite{kardashev}. Finally, as these rates depend on the differential number densities of each species, they depend implicitly on the expansion velocity. Therefore, in order to address the above,
we have developed a numerical code which is based on the one described in \cite{95km}, \cite{97km}. This solves the kinetic equations and  gives the evolution of the distribution of the stable particle populations as a function of the continuously changing comoving source radius.


\section{Model Assumptions} \label{Sec3}
 The picture of an expanding spherical source of radius R(t) carries analogies to the evolution of a Relativistic Blast Wave (RBW) with an initial mass load $M_{0}$ and an initial bulk Lorentz factor $\Gamma_{0}$ \cite{Kazanas02} sweeping mass  from the ambient medium of a wind type density $\rho_{ext}=\rho_{0}~\frac{R_{o}^{2}}{R^{2}}$. Our aim is to expand on the results of a model introduced by  \cite{KaM}, who provided an economic and efficient conversion of the RBW relativistic protons to photons through radiative instabilities akin to that of hadronic supercriticality. Here, however, we have allowed protons to accelerate to ultra high energies, while in case of \cite{KaM} there was no particle acceleration. 

The hot mass is
accumulated with an energy injection rate given by \cite{BM}:
 \begin{equation} \label{eq:4}
          \frac{dE}{dt}=4 \pi R^{2} \rho_{ext} (\Gamma^{2}-\Gamma) c^{3.}
            \end{equation}
 We assume that 90\% of the energy injected above through equation (\ref{eq:4}), is deposited into a Maxwellian distribution of relativistic protons with $E_{p,max}=\Gamma~m_{p}c^{2}$, while the rest  10\% is \mbox producing a high energy power law tail up to  $\gamma_{p,max}>>\Gamma$ \cite{GS}. We also assume that the injection energy rate equals to the total proton luminosity $L_{p}$ inside the source.
 
After sweeping a certain  amount of mass, the RBW starts decelerating and the evolution of M(R) and $\Gamma$(R) is given by the relations \cite{BM}:

    \begin{equation}
        \frac{d \Gamma}{dR} =\frac{-4 \pi R^{2} \rho_{ext} \Gamma^{2}}{M} 
        \end{equation}
          \begin{equation}
         \frac{d M}{dR}=4 \pi R^{2} \rho_{ext} \Gamma
          \end{equation}
             
Under these assumptions, we run the numerical code and search for the onset of supercriticality while the source is expanding at various velosities $u_{exp}$. We define as $L_{p,crit}$ the lowest value of proton luminosity required in order the system to enter to the supercritical regime.
        
\section{Numerical Results}

\subsection{The impact of expansion to the temporal behaviour of supercriticality} \label{ss4}

       As it was mentioned above, the first goal of this work is to investigate whether the expansion of the source has  an impact on the manifestation of the hadronic supercriticalities. In \cite{2012PM}, \cite{2018PM}, it was shown that the temporal behaviour of the system in the supercritical regime is photon outbursts in the form of quasi-periodic oscillations (limit cycles). This is because the system of the kinetic equations in the stationary case bears similarities to the well-known Lotka Volterra equations describing a prey-predator system with predators being the photons and prey the protons.
       
\begin{figure}
\subfloat[]{\includegraphics[width = 3in]{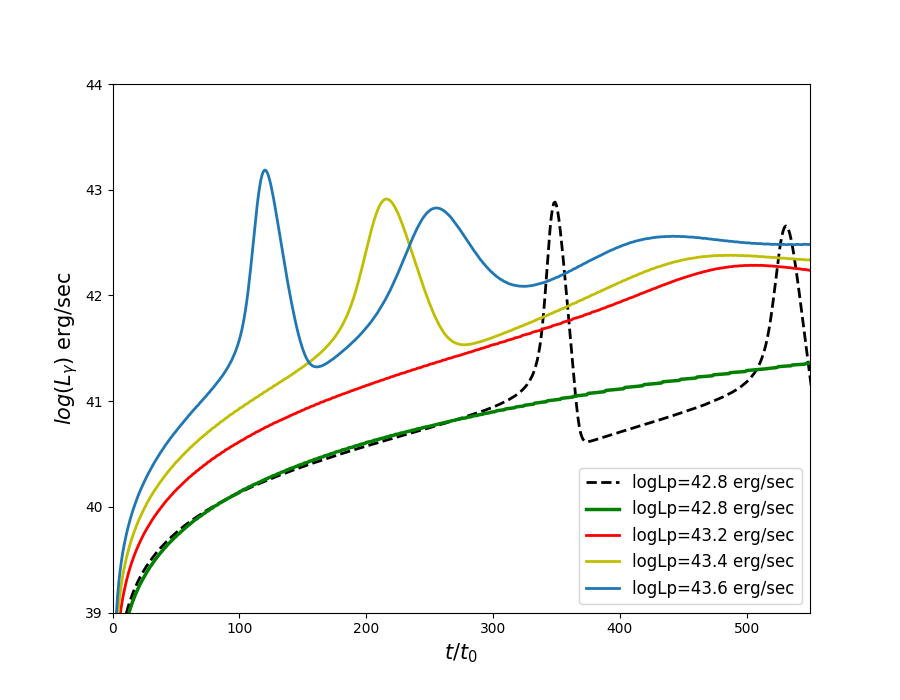}} 
\subfloat[]{\includegraphics[width = 3in]{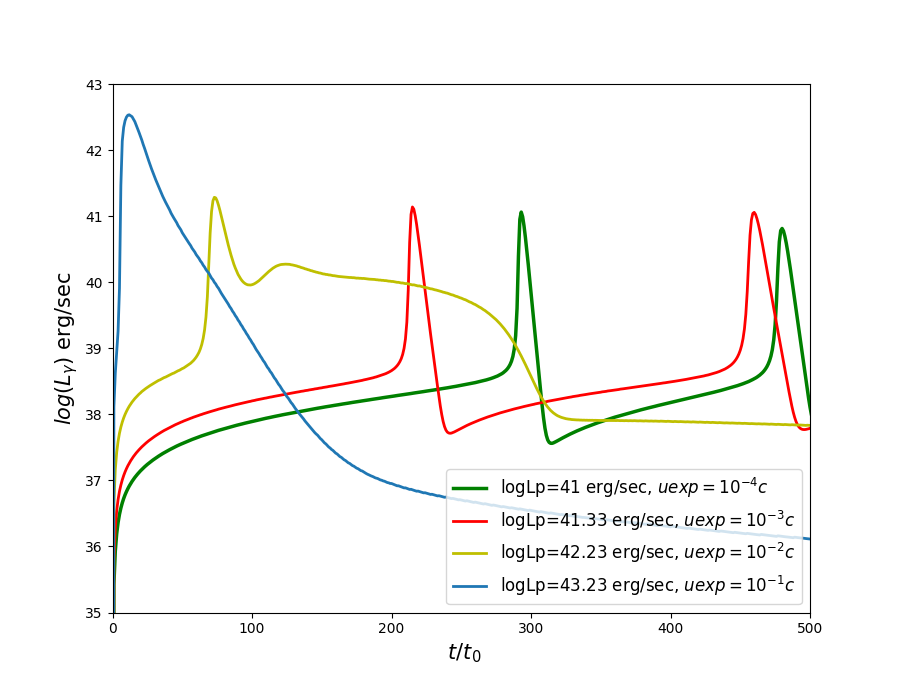}}
\caption{Plots of the photon luminosity versus the ratio $t/t_{0}$ in the comoving frame, where $t_{0}=\frac{R_{0}}{c}$. (a) The black dashed line corresponds to a lightcurve produced by a non expanding source showing limit cycle behaviour for $R_{0}=10^{14}$ cm, $B_{0}=10^3$ G, $\gamma_{max}=10^7$ and $L_{p}=10^{42.8}$ $erg/sec$. The green solid line is the solution provided for the same parameters in case the source is expanding with  $u_{exp}=10^{-2.5}~ c$, $\Gamma_{0}=400$ and $\rho_{0}=10^{-22.8}$ $gr/cm^{3}$. Higher proton luminosities are required in order to make the system again into the supercritical regime (red, yellow and blue solid lines). (b) The green solid line corresponds to a lightcurve produced in the case the source is expanding with a slow expansion velocity ($u_{exp}=10^{-4}c$) for $R_{0}=10^{11}$ cm, $B_{0}=10^4$ G, $\gamma_{max}=10^7$ and $L_{p}=10^{41}$ erg/sec. The other coloured lines correspond to solutions provided for the same parameters in case the source is expanding with higher velocities: $u_{exp}=10^{-3}c$ (red line), $u_{exp}=10^{-2}c$ (yellow line), $u_{exp}=10^{-1}c$ (blue line) where higher critical proton luminosities, $L_{p}$, are required in order to achieve the non linear behaviour.}
\label{fig1}
\end{figure}

       Here we examine whether limit cycle behaviour persists as long as the spherical source is expanding adiabatically. In order to do so we adopt the following procedure: First we search for a set of initial parameters (R, B, $L_{p}$, $\gamma_{max}$) in the stationary case for which the system is in the supercritical regime. Then we let the source expand and we repeat the procedure by varying $L_p$  (or, equivalently, $\Gamma_0$ and $\rho_0$) until we recapture the limit cycle behaviour. For simplicity we have kept B constant. Some characteristic examples can be seen in 
       Figure \ref{fig1}a. The black dashed line depicts a lightcurve produced by a non expanding source showing limit cycle behaviour. However, if we allow a source, having the same initial conditions as before, to expand with some velocity, e.g. $u_{exp}=10^{-2.5}~ c$, the supercriticality disappears completely (green line).  This happens because the relativistic proton column density falls under the critical value as the spherical \mbox volume is increasing. Therefore, one needs higher proton luminosities $L_{p}$ in order to reenter the supercritical regime. Indeed for higher values of $L_{p}$ the supercriticality is recovered, but with a different temporal profile. Thus, by increasing $L_p$ by a factor of 1.6 over its previous value we get a non-linear increase in the photon luminosity but not yet a burst  (red line). Increasing $L_p$ by another factor of 1.6 we obtain one outburst (yellow-green line) and from then on the number of outbursts increases with $L_p$. Evidently the system is once again in the supercritical regime, however the expansion causes  changes in the shape of the outbursts.
       
       
       
       In Figure \ref{fig1}b, each lightcurve corresponds to the onset of supercriticality for different expansion velocities. The green solid line is produced for $u_{exp}=10^{-4}c$. For higher velocities and the same value of proton luminosity, the hadronic supercriticality disappears. The appearance of non linearity features is clearly more energy demanding in these cases. As the  value of $L_{p}$ increases -- here we  increase the value of $\rho_{0}$ while we keep the $\Gamma_{0}$ fixed,  the supercritical behaviour appears again, but at earlier timescales. We also observe different temporal behaviour as the expansion velocity value varies. For low expansion velocities the limit cycle behaviour persists, while for higher expansion velocities, the limit cycles are replaced by lightcurves  which show  fast rise and slow decay.

\subsection{Relevance with the GRB phenomenology}  \label{Sec:4}
\begin{figure} 
\subfloat[]{\includegraphics[width = 3in]{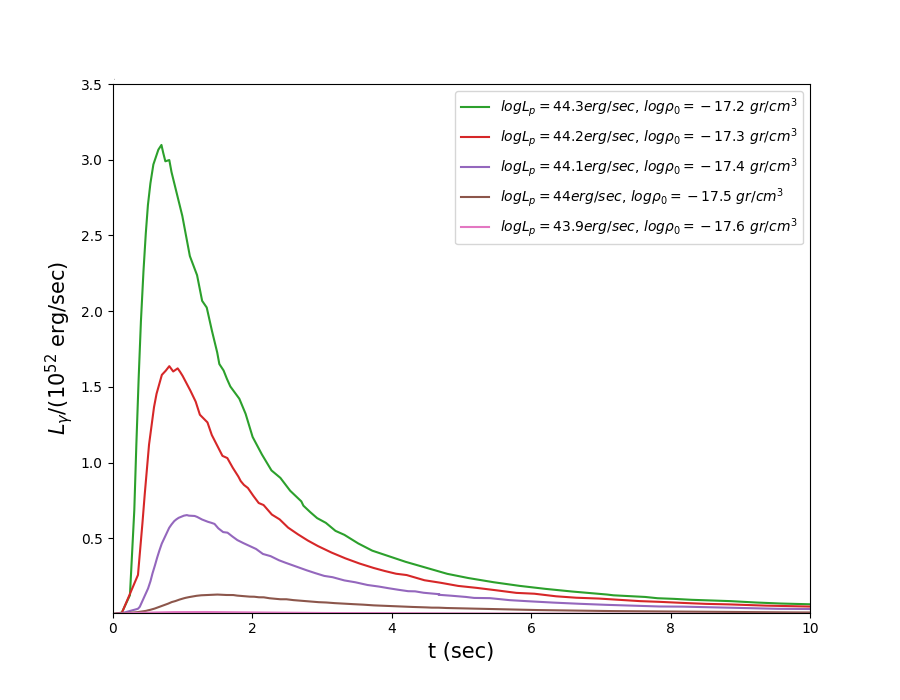}} 
\subfloat[]{\includegraphics[width = 3in]{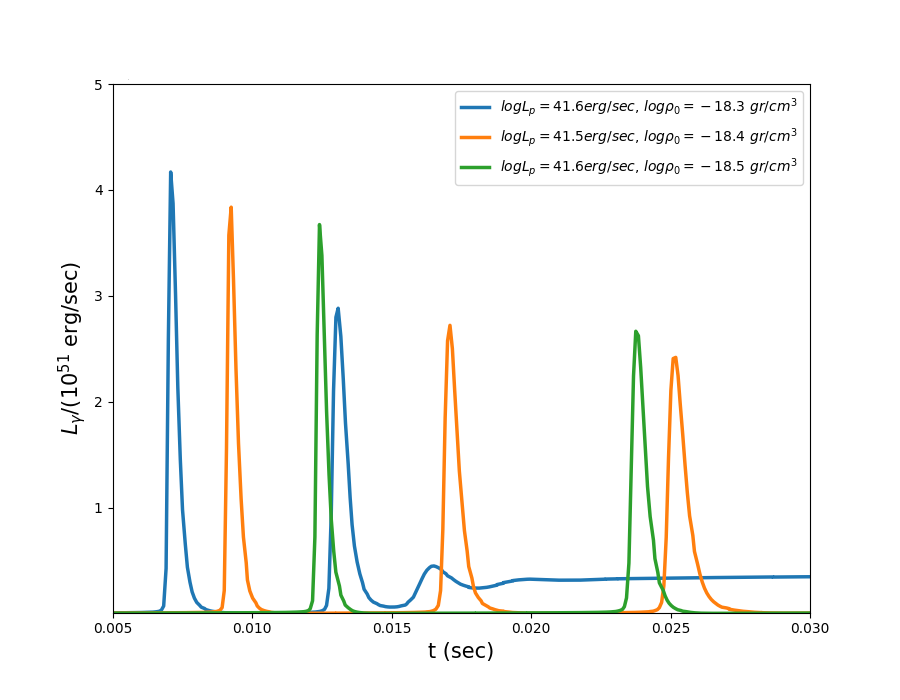}}

\caption{(a) Linear plot of photon luminosity versus time, in the observer's frame, for a sequence of $L_{p}$, in cases where $R_{o}=10^{11}~cm$, $ \Gamma_{0}=400$, $B_{o}=10^{4}~G$ and $u_{exp}=0.1~c$. Supercriticality manifests itself for $L_{p,crit}=10^{43.9}$~ erg/s. (b) Linear plot of photon luminosity versus time for the same parameter values and $u_{exp}=0.001~c$. Supercriticality manifests itself for $L_{p,crit}=10^{41.3}$~ erg/s. }
\label{fig:2}
\end{figure}

As a second goal, we investigate the phase space of initial parameters that lead the system to the supercritical regime and could produce radiation relevant to the GRB prompt emission. For this reason we assume a small spherical source of $R_{0}=10^{11}$ cm, $B_{0}=10^{4}$ G and $\Gamma_{0}=400$, where $R_{0}$,  $B_{0}$, $\Gamma_{0}$ are the initial radius, magnetic field and bulk Lorentz factor respectively. The magnetic field varies as $B \sim R^{-1}$. We examine the source's behaviour in two cases, one for expansion with $u_{exp}=0.1c$ (Fig. \ref{fig:2}a) and as a second case with $u_{exp}=0.001c$ (Fig. \ref{fig:2}b.) .

For such initial parameters the system is led to the supercritical regime for critical proton luminosities, $L_{p}$ shown in the labels of the figures. In case the source is expanding with  $u_{exp}=0.1~c$, the deceleration of the source starts almost instantaneously and the $L_{p}$ is decreased as the source expands. For lower expansion rates ($u_{exp}=0.001~c$) the RBW does not reach the deceleration phase and the injection rate is kept constant.

\begin{figure} 
\center
\includegraphics[width=.6\textwidth]{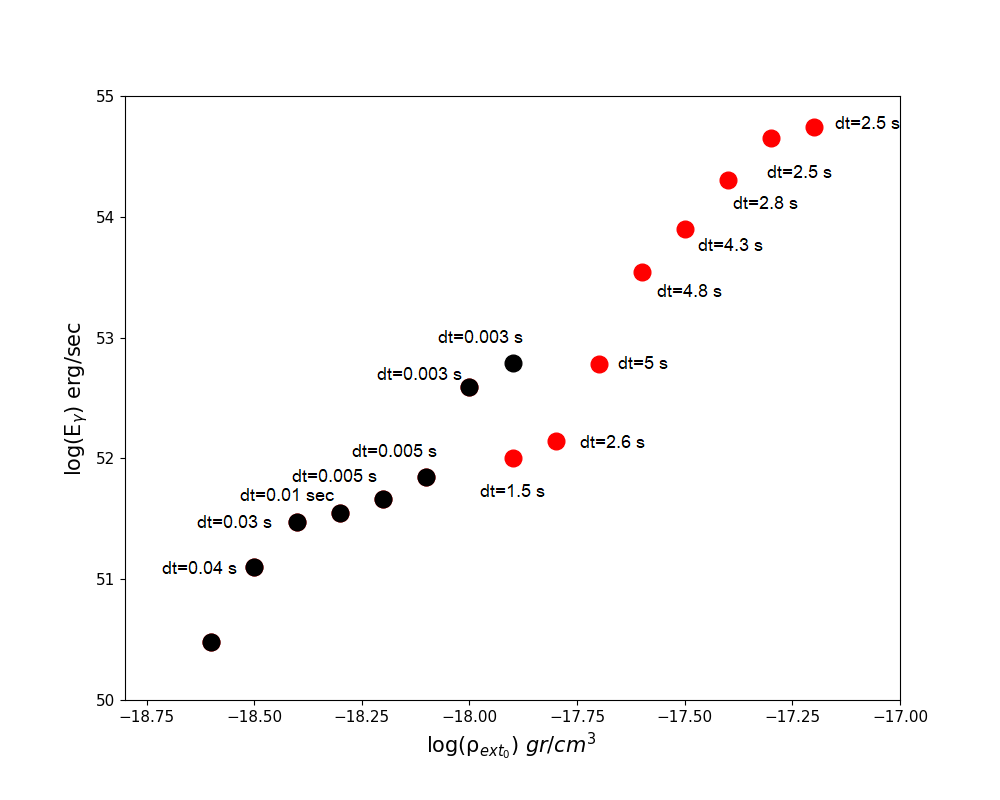}

\caption{The initial density of the ambient medium which provides the  critical proton luminosity $L_{p,crit}$ for the onset of supercriticality versus the bolometric photon energies (which correspond to the lightcurves of Fig. \ref{fig:2}), in the observer's frame. Each point shows also the time duration of the flare, for $u_{exp}=0.1c$ (red points) and  for $u_{exp}=0.001c$ (black points). The first black point has not been annotated with the time duration information because it corresponds to a subcritical solution which does not show limit cycles. }
\label{fig:4}
\end{figure}

The values of initial ambient medium density used in order to produce the proton luminosity $L_{p,crit}$ that leads the system to the supercritical regime  are depicted in the horizontal axis of Fig.\ref{fig:4}. In the vertical axis of this figure we show the bolometric energy released in the observer's frame. Each point of Fig.\ref{fig:4} depicts also the time duration of the flare produced. In the case of slow expansion (Fig.\ref{fig:2}b) the time duration is that of the first spike of each lightcurve and is consistent with the duration of a pulse of a few milliseconds in the GRB prompt emission phase. For higher expansion velocities (Fig.\ref{fig:2}a), the duration of the flare is relevant to that of a long GRB. As far as the energetics are concerned, the photon energy released from each lightcurve in both cases, is compatible to that of $E_{\gamma}=10^{51}-10^{54}$ ergs of the GRB bolometric energy.

\section{Conclusions}
We have presented some preliminary results of a hadronic RBW expanding model which shows supercritical behaviour and presented some first applications to GRBs. We find that for typical values of bulk Lorentz factor, initial radius and magnetic field, this supercritical model can produce bursts of duration and energetics compatible to GRB phenomenology.

As the expansion velocity of the source is increased , the critical proton density required in order to achieve the onset of supercriticality decreases and as a result higher proton luminosities are needed in order to recapture the non linear behaviour. Furthermore, as the expansion velocity varies from low to higher values, the lightcurve's behaviour changes. In Fig.\ref{fig:2} it is clear that sources with faster expansion emit single bursts with high photon luminosity values and a duration of some seconds while sources that expand with lower expansion velocities emit multiple bursts (having a limit cycle behaviour) which are of lower luminosities and each of them shorter in duration (of some milliseconds).

This preliminary work expands on that of \cite{KaM} with the aim to reproduce some of  the results shown there, when the emitting source is expanding adiabatically. A modification in this project is in the \mbox numerical code, which solves the kinetic equations for protons, electrons and photons for sources that are expanding. We have also not taken into account the presence of an upstream medium which scatters the RBW photons. Moreover we allowed a small percentage of the mass accumulated on the RBW to be highly relativistic protons that form a power law distribution. We find that for typical bulk Lorentz factors, this model needs an external medium of a wind type mass densities consistent with those used in \cite{KaM} in order to become supercritical. We note, however, that these values drop if we were to allow a larger fraction of the protons to be in the non-thermal power law tail rather than on the relativistic Maxwellian -- see Section \ref{Sec3}. 

Furthermore, we would like to mention that the value of the critical proton luminosity required in order to insert the supercritical regime will change in case we take into account a different magnetic field profile. In the above analysis we assume that the magnetic field varies as $B \sim R^{-1}$. If $B \sim R^{-2}$, for the same initial parameters, the bursts appear some seconds later and have lower luminosities (about an order of one), compared to those shown in Fig. \ref{fig:2}. One needs slightly higher values of $L_{p}$ in order to recapture the lightcurves shown in Sec. \ref{Sec:4}.

The outlook of this preliminary treatment is that the hadronic RBW model can potentially provide lightcurves with parameters relevant to GRBs.

\section*{Acknowledgments}
 \small{

This research is co-financed by Greece and the European Union (European Social Fund-ESF) through the Operational Programme «Human Resources Development, Education and Lifelong Learning» in the context of the project “Strengthening Human Resources Research Potential via Doctorate Research” (MIS-5000432), implemented by the State Scholarships Foundation (IKY).}

\bibliographystyle{JHEP}
\bibliography{bibliography.bib} 
\label{lastpage}
\end{document}